# Extraterrestrial nucleobases in the Murchison meteorite

Zita Martins[a,b*], Oliver Botta[c,d,1], Marilyn L. Fogel[e], Mark A. Sephton[b], Daniel P. Glavin[c], Jonathan S. Watson[f], Jason P. Dworkin[c], Alan W. Schwartz[g] & Pascale Ehrenfreund[a,c]

[a]*Astrobiology Laboratory, Leiden Institute of Chemistry, 2300 RA Leiden, The Netherlands*

[b]*Department of Earth Science and Engineering, Imperial College, London, SW7 2AZ, UK*

[c]*NASA Goddard Space Flight Center, Code 699, Greenbelt, MD 20771, USA*

[d]*Goddard Earth Sciences and Technology Center, University of Maryland Baltimore County, Baltimore, MD 21228, USA*

[e]*GL, Carnegie Institution of Washington, Washington, DC 20015, USA*

[f]*Planetary and Space Sciences Research Institute, The Open University, Walton Hall, Milton Keynes, MK7 6AA, UK*

[g]*Radboud University Nijmegen, 6525 ED, Nijmegen, The Netherlands*

[1]*Now at International Space Science Institute, Hallerstrasse 6, 3012 Bern, Switzerland.*

*Corresponding author: Zita Martins. Current address: Department of Earth Science and Engineering, Imperial College London, London SW7 2AZ, UK.
Tel: +442075949982. Fax: +442075947444. Email: z.martins@imperial.ac.uk

## Abstract

Carbon-rich meteorites, carbonaceous chondrites, contain many biologically relevant organic molecules and delivered prebiotic material to the young Earth. We present compound-specific carbon isotope data indicating that measured purine and pyrimidine compounds are indigenous components of the Murchison meteorite. Carbon isotope ratios for uracil and xanthine of $\delta^{13}C$ = +44.5‰ and +37.7‰, respectively, indicate a non-terrestrial origin for these compounds. These new results demonstrate that organic compounds, which are components of the genetic code in modern biochemistry, were already present in the early solar system and may have played a key role in life's origin.

# 1. Introduction

One of the most fundamental discoveries of modern science is how nucleic acids store, transcript and translate life's genetic code (Watson and Crick, 1953). Nucleic acids are composed of subunits called nucleotides, each containing a nucleobase, a sugar and a phosphate group. Nucleobases are one-ring (pyrimidines) or two-ring (purines) compounds containing nitrogen atoms. Pyrimidines include uracil, thymine and cytosine, while purines include adenine, hypoxanthine, guanine and xanthine (see section *Appendix A, figure A1* for the structure of nucleobases). Adenine, guanine and cytosine are found in the ribonucleic acid (RNA) and deoxyribonucleic acid (DNA), while thymine is only found in DNA and uracil only in RNA. Hypoxanthine and xanthine are not present in DNA or RNA, but are important intermediates in the synthesis and degradation of purine nucleotides. Since the genetic code is an ancient feature of life, it is likely that Earth's earliest living systems using a genetic code based on the common nucleobases were in existence (Dworkin et al., 2003). One proposed source of these nucleobases is their synthesis on the early Earth by abiotic chemical reactions under plausible primitive Earth conditions, such as the ones applied on the Miller-Urey experiment (Miller, 1953). However, the reducing atmospheric conditions used in this type of experiment are not consistent with the Earth's primitive atmosphere (Kasting, 1993; Kasting and Catling, 2003). Furthermore, it became evident that it is difficult to synthesize prebiotic compounds in a non-reducing atmosphere (Stribling and Miller, 1987). One potentially alternative source of nucleobases is the extraterrestrial delivery of organic material to Earth by comets, asteroids and their fragments as well as interplanetary dust particles (IDPs) (Chyba and Sagan, 1992). Carbonaceous chondrites, which contain many biologically relevant organic compounds (for reviews see e.g. Botta and Bada, 2002; Sephton, 2002) have been analyzed for nucleobases by several different research groups (Hayatsu, 1964; Hayatsu et al., 1968, 1975; Folsome et al., 1971, 1973; Lawless et al., 1972; Van der Velden and Schwartz, 1977; Stoks and Schwartz, 1979, 1981). Hayatsu (1964) reported the detection of N-heterocyclic compounds in an acetylated acid-hydrolyzed Orgueil sample. The nucleobases adenine and guanine, as well as triazines (ammeline and melamine) were identified using paper chromatography. Later, Hayatsu et al. (1968) used HCl-hydrolysis but no acetylation in their analytical procedure to prevent the alteration or destruction of organic compounds during the acetylation step applied in their previous work. Guanylurea and the purine adenine were identified in the Orgueil meteorite, but no guanine was detected. Analysis by a different group (Folsome et al., 1971, 1973; Lawless et al., 1972) using gas chromatography-mass spectrometry (GC-MS) of trimethylsilyl (TMS) derivatives produced clearly different results, with no purines, triazines or guanylurea detected in water and formic acid extracts of the Orgueil, Murchison and Murray meteorites. However, Folsome et al. (1971, 1973) and Lawless et al. (1972) detected 4-hydroxypyrimidine in the same meteorite extracts. Hayatsu et al. (1975) analyzed Murchison samples using the same extraction procedure as Folsome et al. (1971, 1973) and detected only aliphatic amines and alkylpyridines by direct sample volatilisation in a mass spectrometer. However, when more "drastic" extraction conditions were applied (3-6 M HCl or trifluoroacetic acid), Hayatsu et al. (1975) found the purines adenine and guanine, guanylurea and triazines, but no 4-hydroxypyrimidine, and this suggested that purines and triazines were released by acid hydrolysis from the meteorite macromolecular material. Two years later, Van der Velden and Schwartz (1977) analyzed a sample of the Murchison meteorite using high performance liquid chromatography (HPLC) with UV spectroscopy, a technique that did not require derivatization or volatilisation prior to analysis. With this technique xanthine was detected at an abundance of 2.4 parts-per-million (ppm) in formic acid extracts; guanine and hypoxanthine were also tentatively identified (with a concentration of 0.1 ppm and 0.04 ppm, respectively), while no pyrimidines were found at levels higher than the background (0.01 ppm) in water or formic acid extracts. However, after silylation hydroxypyrimidines appeared in the water extracts leading to the suggestion that the compounds detected by Folsome et al. (1971, 1973) and Lawless et al. (1972) might have been produced from contaminants present in the silylation reagents (Van der Velden and Schwartz, 1977). Stoks and Schwartz (1979) re-analyzed water and formic acid extracts of the Murchison, Murray and Orgueil meteorites using specific fractionation techniques (including activated charcoal columns, which adsorb N-heterocyclic compounds, separating the nucleobases from other organic compounds present in the meteorite extract) and ion exclusion chromatography with UV spectroscopy and detected for the first time the nucleobase uracil, in the extracts of all these meteorites. Further re-analysis of the formic acid extracts of Murchison, Murray and Orgueil using GC, HPLC and mass spectrometry (MS) resulted in the detection of xanthine, adenine, hypoxanthine and guanine in concentrations ranging from 114 to 655 parts-per-billion (ppb) in all three meteorites. In contrast, hydroxypyrimidines and triazines were not identified above detection limits of 10 ppb and 50 ppb, respectively, suggesting that previous identifications of triazines by Hayatsu (1964) and Hayatsu et al. (1968, 1975) may have been artefacts synthesized during the experimental procedure. Shimoyama et al. (1990) detected guanine, and possible xanthine and hypoxanthine in the Antarctic meteorite Yamato (Y-) 74662 and Y-791198 meteorites using HPLC with UV spectroscopy. The concentrations of guanine ranged from 30 to 420 ppb, and no pyrimidines were found. Two other Antarctic meteorites, Y-793321 and Belgica (B-) 7904, also yielded no nucleobases (Shimoyama et al. 1990).

Several circumstantial lines of evidence suggested that the origin of the nucleobases present in carbonaceous meteorites is extraterrestrial (e.g. high relative xanthine abundance, low thymine to uracil ratio, and low abundance of cytosine) (Van der Velden and Schwartz, 1977). However, significant quantitative and qualitative variations even between different fragments of the same meteorite (Folsome et al., 1971, 1973; Hayatsu et al., 1975; Van der Velden and Schwartz, 1977; Stoks and Schwatz, 1979, 1981), left open the possibility that terrestrial contamination at the fall site, or during the curation history of the meteorites, as well as analytical artefacts during the extraction, purification and derivatization procedures, could have produced these compounds (Van der Velden and Schwartz, 1977).

Compound specific stable isotope compositions of hydrogen, carbon and nitrogen can be powerful discriminators of the origin of organic compounds in meteorites. For example, the $^{13}$C isotope enrichment of amino acids and carboxylic acids in the Murchison meteorite has been critical to establish the extraterrestrial origin of these compounds (see e.g. Yuen et al., 1984; Engel et al., 1990; Pizzarello et al., 2004; Huang et al., 2005). Accordingly, to establish the origin (terrestrial *vs.* extraterrestrial) of the nucleobases in Murchison, the carbon isotopic ratio of these compounds must be determined. Compound specific isotope measurements of nucleobases in carbonaceous meteorites have not previously been reported.

We subjected the Murchison meteorite (and appropriate controls) to a well-established extraction and isolation procedure (Van der Velden and Schwartz, 1977; Stoks and Schwartz, 1979, 1981) and supplemented it by analyzing the extracts with modern compound-specific carbon isotope ratio instrumentation. Murchison was used in this study to replicate the extraction and isolation procedures used previously (Van der Velden and Schwartz,1977; Stoks and Schwartz, 1979, 1981), and because a relatively large quantity (a few grams) of this meteorite is available. For comparison, a soil sample collected in 1999 in the proximity of the meteorite's 1969 fall site was also subjected to the same extraction, isolation and analytical procedure. To the best of our knowledge, soil from the Murchison meteorite fall site was not collected in 1969. Analysis of soil samples collected in the proximity of meteorite falls were previously shown to be critical in assessing the extent of terrestrial organic contamination in meteorites (e.g. Glavin et al., 1999). In this study, nucleobases were identified in the Murchison meteorite and soil extracts using GC-MS, and their carbon isotope ratios were determined by gas chromatography-combustion-isotope ratio mass spectrometry (GC-C-IRMS).

## 2. Materials and Methods
### *2.1. Extraction and cleaning procedure*

A modification of previously published methods (Van der Velden and Schwartz, 1977; Stoks and Schwartz, 1979, 1981) was applied to our protocol for isolation, extraction and analysis of nucleobases.

An interior piece of about 15g of Murchison meteorite as well as 15g of soil collected near the Murchison recovery location and a serpentine sample (heated to 500ºC for 3 h) used as a procedural blank were separately crushed into powder using a ceramic mortar and pestle. Murchison meteorite powder and soil were placed separately inside Pyrex culture tubes (with Teflon lined screw caps), 1 g per tube. Samples were extracted by ultrasonication with formic acid (8 ml/tube, 3 times) for 1 hour at 60°C. After centrifugation, the acid supernatants were transferred to 15 ml Pyrex tubes and dried under vacuum. Both meteorite and soil formic acid extracts were dissolved in 15 ml 1M HCl, and added separately to columns of 0.6 x 5 cm activated charcoal (charcoal columns were activated as described by Van der Velden and Schwartz, 1977), to which nucleobases are adsorbed. Activated charcoal columns were washed with 1M HCl and $H_2O$ to remove unbound material, and nucleobases were then eluted from the columns with formic acid. The extracts were then dried under vacuum and hydrolyzed (to release the bound fraction of the solvent-soluble nucleobases) as described elsewhere (Stoks and Schwartz, 1979, 1981). The hydrolyzed extracts were diluted with 1M HCl and extracted with ether, followed by charcoal cleaning of the aqueous fraction. The extracts were then dried under vacuum, dissolved in $H_2O$ and subjected to ion-exchange separation with columns (0.4 x 6 cm) of 50W-X8 resin. Uracil and thymine were eluted from these columns with $H_2O$, cytosine and all purines eluted with 5M HCl. Both eluates were dried under vacuum.

The efficiency of the cleaning process was tested by determining the yields of recovery for the different steps involved (charcoal filtration, hydrolysis and ion-exchange separation) using solutions of nucleobases standards of known concentrations. These results as well as the total nucleobase recovery yields (calculated by considering that we applied the charcoal filtration step twice, the hydrolysis step once and the ion-exchange step once) are displayed in the Supplementary Material (see section *Appendix A*). The corresponding technical implications are also discussed in the Supplementary Material (see section *Appendix A*).

All glassware and ceramics used for sample processing were sterilized by annealing in aluminium foil at 500°C for 3 h. Details about chemicals and reagents used in this study are available on-line in the Appendix A.

## 2.2. GC-QMS analysis

The meteorite and soil extracts were dissolved in 500 μl of 0.1N $NH_4OH$ and 30 μl aliquots were dried under vacuum. 10 μl of anhydrous pyridine and 30 μl of BSTFA/TMCS were added to the dried extract residues. Derivatization was carried out at 100°C for 90 min. 2 μl of the resulting solutions were each injected into a GC-QMS (Thermo Finnigan Trace GC coupled to a Thermo Finnigan Trace DSQ QMS). Due to the lack of GC columns optimized for nucleobase compounds, various GC operating conditions were tested to optimize the peak shape of nucleobases, including different GC columns, temperature programs and carrier gas flow rates. Optimized conditions are as follow. Splitless injection with He as carrier gas at a constant pressure of 13 PSI was used. Separation was performed on a HP Ultra 2 (25 m x 0.32 mm ID x 0.17 μm film thickness) column. The GC oven temperature was held for 1 min at 75°C and ramped to 300°C at a rate of 5°C $min^{-1}$ and then held for 5 min. The presence of nucleobases was confirmed by retention time comparison to standards and by their unique mass fragmentation pattern.

## 2.3. GC-C-IRMS analysis

300 μl aliquots of the meteorite and soil extracts (out of the 500 μl) were carried through the same procedure as described for GC-QMS analysis. Carbon isotope analyses were performed using a Thermo Finnigan MAT Delta Plus XL GC-C-IRMS. Temperature program, carrier gas and pressure were the same as the GC-QMS analysis. The GC column in the GC-QMS was removed and then installed in the GC-C-IRMS. Compounds separated by the GC column were converted to $CO_2$ through an oxidation oven kept at 980ºC. $CO_2$ reference gas with a known $\delta^{13}C$ value (-41.10‰ PDB) was injected via the interface to the IRMS, for the computation of $\delta^{13}C$ values of sample and standard compounds. Peaks corresponding to the compounds of interest were integrated using the software supplied with the GC-C-IRMS instrument, which corrects for background, calculates and reports $\delta^{13}C$ values. Standards for the analyses included pyrene, with a $\delta^{13}C$ value of -24.03‰ (±0.16‰) when measured by GC-C-IRMS. Additionally, individual nucleobases standards were subjected to the entire derivatization procedure described above and run on the GC-C-IRMS, with typical standard deviation of ±0.44‰. Corrections for carbon added from the BSFTA were calculated by mass balance: $\delta^{13}C$ nucleobase in sample derivatized = (% of carbon nucleobase) ($\delta^{13}C$ nucleobase sample) + (% of carbon BSTFA) ($\delta^{13}C$ BSTFA). The average $\delta^{13}C$ values used for BSTFA were -48.99‰ ± 0.1‰ (from uracil standards), -44.91‰ ± 0.38‰ (from thymine standards), and -40.47‰ ± 0.34‰ (from xanthine standards), and were obtained by mass balance: $\delta^{13}C$ nucleobase standard derivatized = (% of carbon nucleobase) (EA nucleobase standard) + (% of carbon BSTFA) ($\delta^{13}C$ BSTFA), where the EA nucleobase standard value corresponds to the $\delta^{13}C$ value of the nucleobase standard established by a Carlo Erba elemental analyzer (EA)-IRMS with He as the carrier gas. The uncertainties in the $\delta^{13}C$ values ($\delta x$) are based on the standard deviation of the average value of between three and four separate measurements (N) with a standard error $\delta x = \sigma_x \cdot N^{-1/2}$.

## 3. Results

Following the literature protocol (Van der Velden and Schwartz, 1977; Stoks and Schwartz, 1979, 1981), an interior fragment of the Murchison meteorite was extracted and purified for nucleobase isotopic analyses. This procedure substantially limited the presence of interfering compounds and was optimized for the detection of uracil and xanthine (see section *2. Materials and Methods* for details). A detailed study of the yields of recovery for each cleaning step during the purification process is described in section *Appendix A*. A relatively large quantity of Murchison meteorite (15 g) was necessary to perform carbon isotope measurements of nucleobases (see section *2. Materials and Methods*). The limit of detection of the GC-C-IRMS, combined with the limited mass availability of meteorite samples, prevented us from performing stable nitrogen or hydrogen measurements of the nucleobases.

Two chromatographic traces obtained from the GC-C-IRMS analysis of the water eluate from the ion-exchange separation of the formic acid extract of the Murchison meteorite are shown in Figures 1a : the *m/z* 44 ($^{12}CO_2$) trace (bottom) and the ratio *m/z* 45/44 ($^{13}CO_2/^{12}CO_2$) (top). These traces include the peak corresponding to BSTFA-derivatized uracil, assigned by retention time comparison with BSTFA-derivatized authentic uracil standard analyzed on the same instrument. Confirmation of this assignment was achieved by comparison of the mass fragmentation patterns of the corresponding peak in the meteorite extracts (Figure 1b) with the mass spectra of a BSTFA-derivatized authentic uracil standard (Figure 1b inset), analyzed by gas chromatography-quadrupole mass spectrometry (GC-QMS) using the same GC column and analytical conditions that were used for the GC-C-IRMS measurements. The same analysis was performed for the hydrochloric acid eluate from the ion-exchange separation. The two traces *m/z* 44 (bottom) and ratio *m/z* 45/44 (top) that include the peak corresponding to BSTFA-derivatized xanthine are shown in Figure 2a, and the corresponding GC-QMS mass fragmentation patterns of the peaks in the meteorite extract and the authentic xanthine standard are shown in Figure 2b. The peak shapes in the *m/z* 45/44 traces (top in Figures 1a and 2a) do not correspond to the typical sinusoidal *m/z* 45/44 traces, in which *m/z* 45 goes through the GC column slightly ahead of *m/z* 44. This can be explained by the lack of GC-columns optimized for nucleobase compounds (see section *2.2. GC-QMS analysis*), and in particular for isotope measurements, as the *m/z* 45/44 traces of BSTFA-derivatized authentic nucleobase standards analyzed under the same conditions showed the same behavior.

## 4. Discussion

### *4.1. Compound-specific carbon isotopic measurements*

Analysis of the data from the GC-C-IRMS measurements yielded $\delta^{13}C$ values of +44.5‰ (± 2.3‰) for uracil and +37.7‰ (± 1.6‰) for xanthine in the Murchison meteorite (Table 1). These values fall within the range of those measured for extraterrestrial amino acids and carboxylic acids in carbonaceous chondrites (Yuen et al., 1984; Engel et al., 1990; Pizzarello et al., 2004; Huang et al., 2005). In order to constrain the possible contributions of terrestrial nucleobases to the carbon isotope values for uracil and xanthine measured in Murchison, GC-C-IRMS analyses were also carried out for the nucleobases present in the Murchison soil extract (Table 1). Soil uracil has a $\delta^{13}C$ value of -10.6‰ (±1.8‰) and xanthine was below the detection limit of GC-C-IRMS (~1 ppb). Thus, there should be no terrestrial contribution from the landing site soil to the value for xanthine measured in the meteorite. For uracil, any terrestrial contamination from the soil would decrease the measured $\delta^{13}C$ value in the Murchison meteorite extract. While our analytical methods were not optimized for the detection of other nucleobases, we were able to detect thymine in the soil ($\delta^{13}C$ = -15.9‰ ± 1.1‰). The negative $\delta^{13}C$ values measured for uracil and thymine in the soil are in the range expected for terrestrial organic compounds of biological origin (for review see e.g. Sephton and Botta, 2005; Scott et al., 2006) and are clearly distinct from the positive $\delta^{13}C$ values of uracil and xanthine we have measured in the Murchison meteorite.

Extraterrestrial dicarboxylic acids are the most abundant class of compounds detected in the Murchison meteorite extracts. Their measured $\delta^{13}C$ values were in the range of +28‰ to +44‰ (see Table 2 and section *4.2. Carboxylic acids in the Murchison meteorite and soil samples*), consistent with previous results (Pizzarello and Huang, 2002). These compounds were chromatographically separated from the nucleobases (different retention time) and therefore did not interfere with our carbon isotope measurements. Comparison of the mass fragmentation patterns of the meteorite extracts to standards indicates the possible presence of co-eluting compounds with the BSTFA-derivatized xanthine peaks. Of these, the most conspicuous interfering peak is the m/z 313 present in the Murchison xanthine spectrum (Figure 2b). This mass fragment corresponds to BSTFA-derivatized hexadecanoic acid (a monocarboxylic acid), which is also observed at the same retention time in the soil extract, and has m/z 313 (Figure 3). Thus, it is very likely that this compound in the meteorite has a terrestrial origin and carries a light isotopic signature. In addition, the high carbon number of hexadecanoic acid ($C_{16}$) is inconsistent with known extraterrestrial meteoritic monocarboxylic acids that range from $C_2$ up to $C_{12}$ (Naraoka et al., 1999; Huang et al., 2005 and references therein).

A background of unresolved compounds, which co-elute with the meteoritic nucleobases is observed in Figures 1 and 2. For example, the characteristic fragments of BSTFA-derivatized xanthine present in the meteorite extract are clearly evident in the GC-QMS mass spectrum (Figure 2b). In addition to this, there is a continuum of ions derived from the chromatographically unresolved background material. Software supplied with the GC-C-IRMS instrument corrects for this background and therefore it will not interfere with the reported $\delta^{13}C$ values for uracil and xanthine present in the Murchison meteorite.

Based on these arguments, the measured carbon isotope values for uracil (possible contribution of terrestrial uracil from the soil) and xanthine (possible co-elution of terrestrial hexadecanoic acid) in the Murchison meteorite should be considered to be lower limits. Given the high positive $\delta^{13}C$ values for uracil and xanthine measured for Murchison meteorite extracts, these interferences do not compromise the conclusion that these two nucleobases are definitely of extraterrestrial origin.

### 4.2. Carboxylic acids in the Murchison meteorite and soil samples

The identification (by retention time and mass fragmentation patterns) of several peaks in the GC-QMS total ion current (TIC) was essential to determine whether the nucleobase peak assignments (both in the Murchison meteorite and in the soil) were correct and if the nucleobase peaks were separate and distinct from other compounds.

The detection and carbon isotope compositions have already been published for most of the dicarboxylic acids present in the Murchison meteorite (Lawless et al., 1974; Peltzer et al., 1984; Cronin et al., 1993; Pizzarello and Huang, 2002). Despite the extensive fractionation procedure applied to isolate nucleobases from other compounds in the meteorite and soil extracts, GC-QMS analyses show that dicarboxylic acids are still present in the purified formic acid extracts of both the Murchison meteorite and soil. Since dicarboxylic acids are present in both the $H_2O$ and HCl eluates of the Murchison meteorite (Figures 1a and 2a), they might be trailing in the ion exchange separation step (see section *2.2. Extraction and cleaning procedure*). This could be due to overloading of the ion-exchange columns as well as the presence of other compounds which would cause elution of the dicarboxylic acids. It is not clear if additional purification steps would have removed these interferences, but it would definitely have led to further sample loss preventing the stable carbon isotope measurements of nucleobases. The presence of dicarboxylic acids in the eluates did not interfere with the isotopic analysis of the nucleobases since they were chromatographically separated on the GC-C-IRMS column. Dicarboxylic acids present on the GC-C-IRMS traces for both the $H_2O$ eluate (Figure 1a) and the HCl eluate of the Murchison meteorite (Figure 2a) have been identified by GC-QMS as butanedioic acid, 2-methyl-butanedioic acid, 2,3-dimethylbutanedioic acid, pentanedioic acid, 2-methylpentanedioic acid, 3-methylpentanedioic acid, hexanedioic acid, and 1,2-benzenedicarboxylic acid (peaks 2, 3, 5 to 8, 11 and 13 in the $H_2O$ eluate, Figure 1a; peaks 1 to 8 in the HCl eluate, Figure 2a). Stable carbon isotope values of dicarboxylic acids present in the Murchison meteorite obtained in a previous study range from +19.1‰ to +28.1‰ (Pizzarello and Huang, 2002). The $\delta^{13}C$ values for the Murchison meteorite dicarboxylic acids measured in this study are in agreement with these literature values, or are slightly higher (Table 2). The only exception is pentanedioic acid, whose $\delta^{13}C$ value of +44.0‰ is significantly higher than the $\delta^{13}C$ value of +26.8‰ published previously for Murchison (Pizzarello and Huang, 2002). The difference between the two measurements could be due to a higher degree of terrestrial contamination in the sample of Murchison from the previous measurement, since pentanedioic acid is a common terrestrial contaminant found in the biosphere (Pizzarello and Huang, 2002), leading to a small decrease in the $\delta^{13}C$ value. We cannot exclude the possibility of a small amount of an isotopically heavy compound co-eluting with pentanedioic acid in our analysis, which would increase the carbon isotope value for pentanedioic acid.

### *4.3. Origin of meteoritic nucleobases*

It is generally accepted that extraterrestrial nucleobases could have been formed by abiotic reaction mechanisms in a variety of cosmic environments. However, a low formation rate combined with a low stability against UV radiation makes the detection of nucleobases in the interstellar and circumstellar medium extremely difficult (Peeters et al., 2003). In fact, only upper limits of this class of compounds were detected in the interstellar medium (Kuan et al., 2003). Instead, synthetic processes on the meteorite parent body during aqueous alteration are more likely to be responsible for the presence of meteoritic nucleobases. A number of abiotic synthetic routes have been investigated in laboratory simulations. These include the polymerization of hydrogen cyanide (HCN) (Oró, 1960, 1961; Oró and Kimball 1961; Sanchez et al., 1967; Ferris et al., 1978; Voet and Schwartz, 1983; Schwartz and Bakker, 1989; Minard et al., 1998; Levy et al., 1999; Miyakawa et al., 2002), synthesis by quenching a $CO-N_2-H_2O$ high-temperature plasma (Miyakawa et al., 2000), the reaction of cyanoacetylene with cyanate in relative dilute solution at pH 8 and room temperature (Ferris et al., 1968), and the reaction of cyanoacetaldehyde with urea in eutectic solution (Nelson et al., 2001) or at higher temperature (Robertson and Miller, 1995). Other pathways are obviously also possible (for an overview see Ferris and Hagan, 1984; Orgel, 2004), and a number of them might have occurred on the Murchison meteorite parent body. Degradation of nucleobases in the hydrated parent body environment also has to be considered. For example, cytosine degrades to uracil with a half-life of 17,000 years and guanine decomposes to xanthine with a half-life of 1.3 Ma (Levy and Miller, 1998) at 0ºC and pH 7. Consequently, meteoritic nucleobase distributions are the result of both synthetic and subsequent degradation reactions.

## 5. Conclusions

By demonstrating that one purine and one pyrimidine in the Murchison meteorite are extraterrestrial in origin, a large variety of the key component classes in terrestrial biochemistry, including amino acids, sugar related compounds (Cooper et al., 2001), carboxylic acids and nucleobases, have been identified as indigenous components in the Murchison meteorite (for a review see e.g. Botta and Bada, 2002; Sephton, 2002). Our data advance proposals that life's raw materials were delivered to the early Earth and other planetary bodies by exogenous sources, including carbonaceous meteorites. In contrast, the endogenous synthesis of prebiotic organic compounds may have been constrained by the conditions on the young Earth, perhaps most importantly by the oxidation state of the atmosphere. For example, only low yields of amino acids were produced under non-reducing conditions in the Miller-Urey-type experiment (Stribling and Miller, 1987). Yet, whatever the inventory of endogenous organic compounds on the ancient Earth, it would have been augmented by extraterrestrial material. It is estimated that these sources delivered ~$10^9$ kg of carbon per year to the Earth during the heavy bombardment phase 4.5–3.9 billion years ago (Chyba and Sagan, 1992).

In modern biology uracil is ubiquitous as a nucleobase in RNA, while the role of xanthine is limited in modern biochemistry (Kulikowska et al., 2004), most notably as an intermediate in the biosynthesis of guanosine and uric acid. It is also interesting to note that both xanthine and uracil are capable of self-association in monolayers, which might have been of importance in prebiotic chemistry on mineral surfaces on the early Earth (Sowerby and Petersen, 1999). A continuous influx of meteoritic uracil and xanthine and possibly other nucleobases would have enriched the prebiotic organic inventory necessary for life to assemble on the early Earth. Following the birth of the Solar System, carbonaceous meteorite infall would have been common on all terrestrial planets. Consequently, nucleobases delivered to these worlds together with sugar-related species and amino acids might have been beneficial to the origin of life on Earth, Mars, or elsewhere.

## Acknowledgements

We are grateful to R. D. van der Hilst and two anonymous reviewers for their helpful comments. This project was supported by Fundação para a Ciência e a Tecnologia (scholarship SFRH/BD/10518/2002), NASA Astrobiology Institute, the NASA Exobiology and Evolutionary Biology Program and through cooperative agreement NNA04CC09A, Goddard Center for Astrobiology/NASA Astrobiology, European Space Agency, NWO-VI 016023003 and PPARC. The authors would like to thank L. Welzenbach (Smithsonian National Museum of Natural History, Washington DC) for providing us with a Murchison sample and Z. Peeters for graphic support.

## Appendix A. Supplementary data

Supplementary data associated with this article can be found in the online version.

# Figure Captions

**Figure 1 - GC-C-IRMS analysis of the BSTFA-derivatized formic acid extract, water eluate of the Murchison meteorite.** (a) The *m/z* 44 ($^{12}CO_2$) trace (bottom) and the ratio between the *m/z* 45 and *m/z* 44 ($^{13}CO_2/^{12}CO_2$) trace (top) for the GC-C-IRMS analysis are displayed. The insets show the uracil region of each chromatogram. The following peaks were tentatively identified by GC-QMS: 1. 2-hydroxyhexanoic acid; 2. butanedioic acid; 3. 2-methylbutanedioic acid; 4. unidentified; 5. 2, 3-dimethylbutanedioic acid; 6. pentanedioic acid; 7. 2-methylpentanedioic acid; 8. 3-methylpentanedioic acid; 9. 3-ethylpentanedioic acid; 10. ethylpentanedioic acid; 11. hexanedioic acid; 12. heptanedioic acid; 13. 1,2-benzenedicarboxylic acid; 14. unidentified; 15. unidentified. (b) The GC-QMS mass spectrum for the peak assigned to BSTFA-derivatized uracil and its structure. The inset shows the mass spectrum of a BSTFA-derivatized uracil standard.

**Figure 2 - GC-C-IRMS analysis of the BSTFA-derivatized formic acid extract, hydrochloric acid eluate of the Murchison meteorite.** (a) The *m/z* 44 ($^{12}CO_2$) trace (bottom) and the ratio between the *m/z* 45 and *m/z* 44 ($^{13}CO_2/^{12}CO_2$) trace (top) for the GC-C-IRMS analysis are displayed. The insets show the xanthine region of each chromatogram. The following peaks were tentatively identified by GC-QMS: 1. butanedioic acid; 2. 2-methylbutanedioic acid; 3. 2, 3-dimethylbutanedioic acid; 4. pentanedioic acid; 5. 2-methylpentanedioic acid; 6. 3-methylpentanedioic acid; 7. unidentified; 8. 1,2-benzenedicarboxylic acid; 9. unidentified; 10. unidentified. (b) The GC-QMS mass spectrum for the peak assigned to BSTFA-derivatized xanthine and its structure. The inset shows the mass spectrum of a BSTFA-derivatized xanthine standard.

**Figure 3 - GC-QMS mass spectrum corresponding to hexadecanoic acid present in the Murchison soil.** The structure of BSTFA-derivatized hexadecanoic acid is shown.

Figure 1 Martins et al.

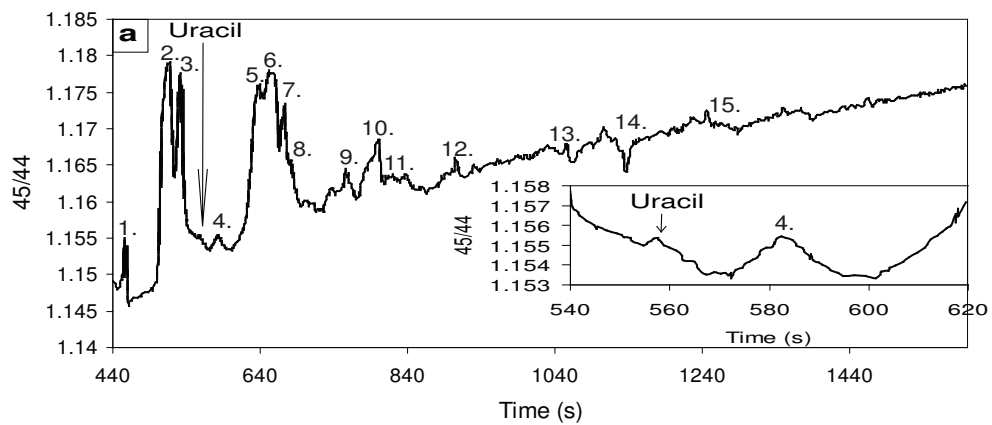

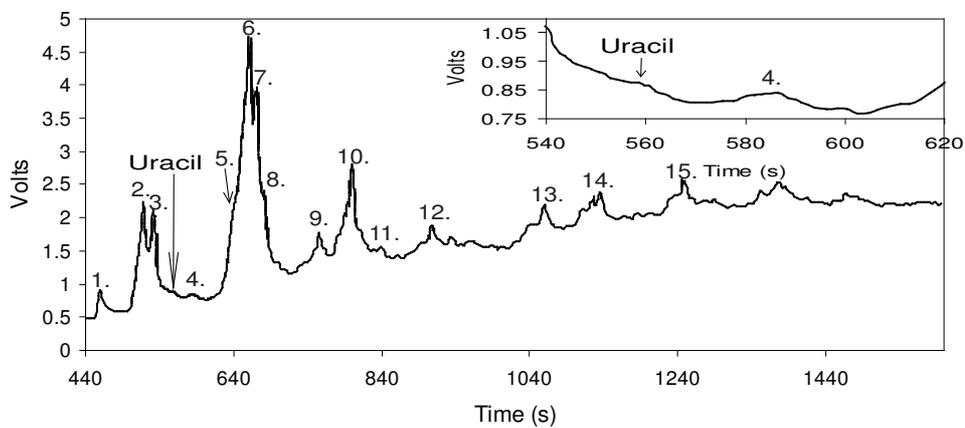

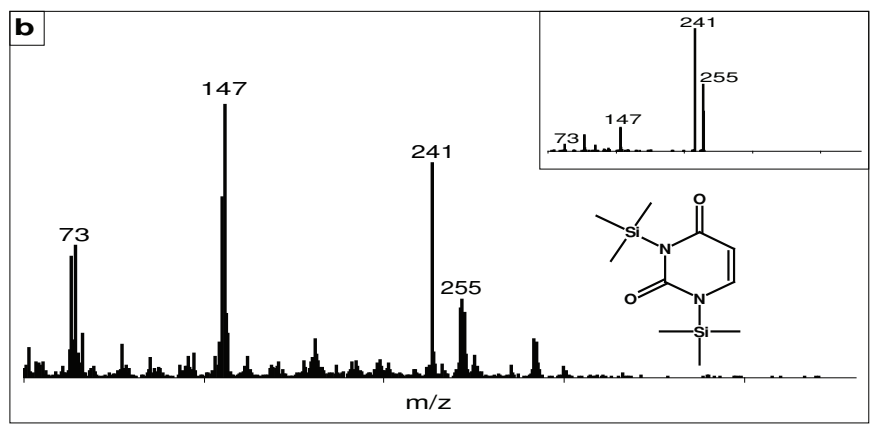

Figure 2 Martins et al.

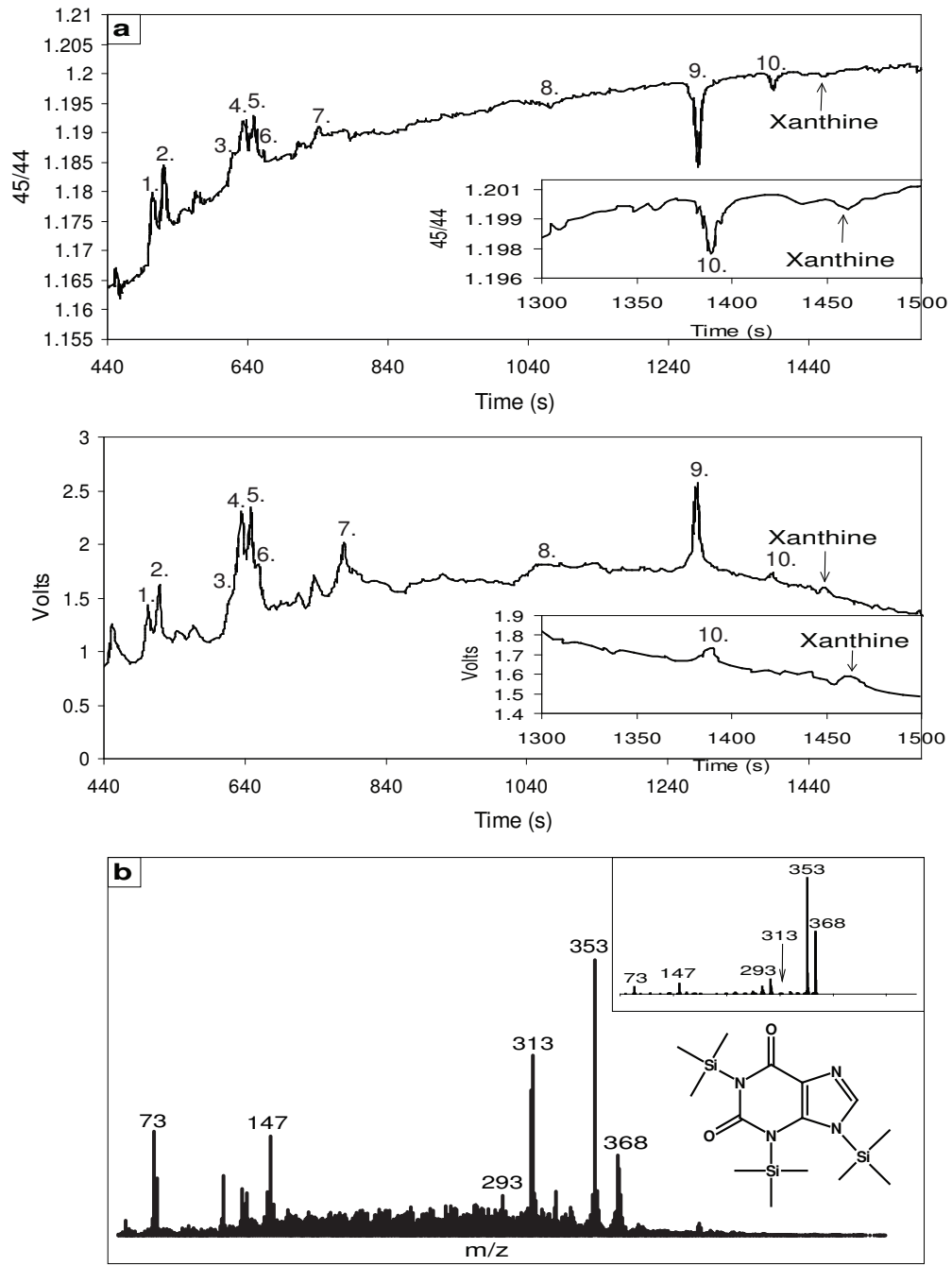

Figure 3 Martins et al.

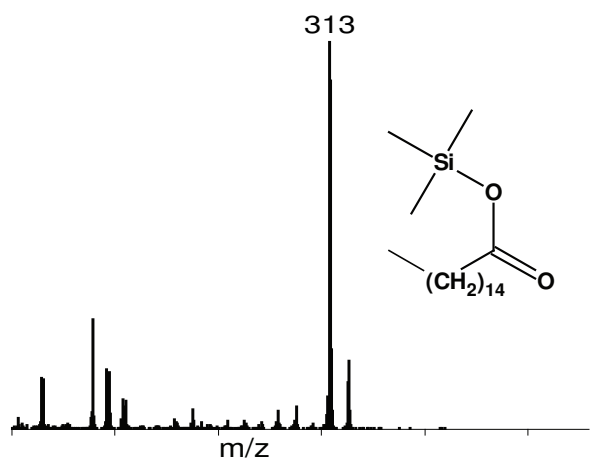

**Table 1 -** $\delta^{13}C$ **values (‰) of nucleobases in the Murchison meteorite and soil samples.**

| $\delta^{13}C$ | Uracil | Xanthine | Thymine |
|---|---|---|---|
| Murchison meteorite | +44.5 ± 2.3 | +37.7 ± 1.6 | n.d. |
| Soil | -10.6 ± 1.8 | n.d. | -15.9 ± 1.1 |

n.d. - not determined due to low concentrations.

**Table 2** - $\delta^{13}C$ values (‰) of dicarboxylic acids in the Murchison meteorite.

| Dicarboxylic acids | This study | Literature* |
|---|---|---|
| butanedioic acid | +30.1 | +28.1 |
| 2-methylbutanedioic acid | +28.0 | +26.5 |
| pentanedioic acid | +44.0 | +26.8 |
| 2-methylpentanedioic acid | +34.2 | +27.9 |
| 3-methylpentanedioic acid | +28.0 | +19.1 |
| hexanedioic acid | +28.4 | +21.4 |

*Pizzarello and Huang (2002).

# Appendix A

## A1. Materials and Methods

### A1.1 Chemicals and reagents

Purine and pyrimidine standards comprised cytosine, uracil, hypoxanthine, guanine, xanthine, thymine (Sigma-Aldrich, ≥ 97% purity) and adenine (Merck, ≥ 99% purity). A stock solution of these nucleobases was prepared by dissolving each nucleobase in 0.1N ammonium hydroxide. Ammonium formate, ammonium hydroxide, HPLC grade water, bis(trimethylsilyl)-trifluoroacetamide with 1% trimethylchlorosilane (BSTFA/TMCS), and anhydrous pyridine were purchased from Sigma-Aldrich. Formic acid p.a. (assay ≥ 99%) was purchased from Acros Organics, HPLC-S gradient grade acetonitrile from Biosolve Ltd., hydrochloric acid (37%) from Merck, and diethyl ether from Riedel-de-Haën. AG® 50W-X8 cation exchange resin (100-200 mesh) was purchased from Bio-Rad and activated cocoanut charcoal (50-200 mesh) from Fisher. Biopur tips (Fisherbrand) and Eppendorf® microcentrifuge tubes (Sigma-Aldrich) were supplied sterilized.

### A1.2 Determination of the nucleobase recoveries

The efficiency of the cleaning process was tested by determining the yields of recovery for the different steps involved (charcoal filtration, hydrolysis and ion-exchange separation) using solutions of nucleobases standards of known concentrations. These results as well as the total nucleobase recovery yields (calculated by considering that we applied the charcoal filtration step twice, the hydrolysis step once and the ion-exchange step once) are displayed in TableA1. The corresponding technical implications are discussed in section *A2 Technical implications*.

#### A1.2.1 Charcoal filtration

The recoveries were determined by passing mixtures of nucleobase standards (5 μg of guanine and 4 μg each of uracil, thymine, xanthine, hypoxanthine and adenine, and a second mixture of nucleobase standards ten times more diluted) in 5 ml of 1M HCl through columns of 0.6 x 5 cm activated charcoal beds. These columns were washed with 5 ml 1M HCl and 5 ml $H_2O$ each. The charcoal columns were first eluted with 5 ml of formic acid, and an additional elution of 5 ml of formic acid was collected.

All collected fractions were dried under vacuum, brought up in 0.1N NH$_4$OH and analyzed by HPLC-UVS (see section *A1.3* for HPLC-UV analysis).

*A1.2.2 Acid hydrolysis*

To test the hydrolysis step, individual nucleobase standards solutions were placed separately in different test tubes and then dried under vacuum. The amounts of nucleobases in the test tubes were 4 µg each of guanine and xanthine, and 5 µg of cytosine, uracil, thymine, hypoxanthine and adenine. The samples were hydrolyzed with 8 ml 3M HCl, at 110°C for 18 h. Subsequently, samples were dried, brought up in 0.1N NH$_4$OH and analyzed by HPLC-UVS (see section *A1.3* for HPLC-UV analysis).

*A1.2.3 Ion-exchange separation*

The recoveries were determined by loading separate 50W-X8 resin columns (0.4 x 6 cm) each with 5 ml H$_2$O solutions containing 4 µg of guanine and xanthine, and 5 µg each of cytosine, uracil, thymine, hypoxanthine and adenine. The columns were first eluted two times with 5 ml of H$_2$O, then two times with 5 ml of 5M HCl. All eluates were dried, re-dissolved in 0.1N NH$_4$OH and analyzed for purines and pyrimidines by HPLC-UVS (see section *A1.3* for HPLC-UV analysis).

*A1.3 HPLC-UVS analysis*

HPLC separation was achieved using a dC$_{18}$ reverse phase Atlantis$^{TM}$ column (Waters®, 4.6 x 150 mm, 5 µm), at room temperature and a flow rate of 1 ml min$^{-1}$. Buffer A was 10 mM ammonium formate buffer and buffer B was acetonitrile. The gradient used was 0 to 0.01 min., 0% buffer B; 0.01 to 15 min., 0 to 3% buffer B; 15 to 15.01 min., 3 to 0% buffer B; 15.01 to 25 min., 0 % buffer B. UV absorption was monitored with diode array detection on a Shimadzu SPD-M10A$_{VP}$, (spectral acquisition range from 190 nm to 370 nm). Peaks were integrated using the maximum absorbance wavelength for each nucleobase: 257 nm for uracil, 248 nm for hypoxanthine, 246 nm for guanine, 265 nm for xanthine, 263 nm for thymine and 259 nm for adenine. Nucleobases were identified by comparison of the retention times of each peak at the respective wavelength with those of a standard.

## A2. Technical implications

Preliminary analysis of formic acid extracts of Murchison by HPLC-UVS showed the presence of underlying UV absorbing material, namely abundant aliphatic and aromatic carboxylic acids (identified by GC-MS), which interfered with the detection of the nucleobases. For this reason an isolation process subsequent to the formic acid extraction was performed based on the method used by Van der Velden and Schwartz (1977) and Stoks and Schwartz (1979, 1981).

Minimal amounts of solvents were used in the isolation steps in order to minimize sample loss and contamination. Results show that it is necessary to use 10 ml of formic acid to completely elute 5 μg of guanine and 4 μg of uracil, thymine, xanthine, hypoxanthine and adenine from the charcoal columns. The overall average recovery (Table A1) of the two solutions, one containing 5 μg of guanine and 4 μg of all the other nucleobases, and a solution diluted by a factor of ten, was 71% (values ranging from 47% for guanine to 97% for hypoxanthine). Our value is identical to that of Stoks and Schwartz (1981), who found an average nucleobases recovery for the formic acid extract after charcoal cleaning of 71%. Stoks and Schwartz (1979, 1981) used much larger volumes of solvent, which led to a more time consuming process with no apparent gain in the nucleobases recovery.

The average nucleobase recovery for acid hydrolysis was 88% (Table A1), with values for each nucleobase of 81% for thymine, 82% for xanthine, 83% for adenine (hydrolysed to hypoxanthine), 85% for cytosine (hydrolysed to uracil), 95% for guanine (hydrolysed to xanthine) and hypoxanthine, and 96% for uracil. Our average recovery of purines and pyrimidines after 3 M HCl hydrolysis is substantially better than the 60% obtained by Stoks and Schwartz (1981).

Our analysis revealed that 5 ml of $H_2O$ was necessary to elute 5 μg of uracil and 5 μg of thymine from the 50W-X8 resin, and that for all other nucleobases (4 μg of guanine and xanthine, and 5 μg each of cytosine, hypoxanthine and adenine) 10 ml of 5 M HCl was required. The overall average recovery for the ion-exchange separation step was 72% (Table A1) with values ranging from 48% for xanthine to 96% for cytosine.

The overall recovery of nucleobases estimated from these individual steps was 32% (see Table A1).

# Figure Captions

**Figure A1 – Structures of nucleobases.** Nucleobases are one-ring (pyrimidines) or two-ring (purines) N-heterocyclic compounds. The structure of purines (two-ring N-heterocyclic compounds) is shown on the top, while the structure of pyrimidines (one-ring N-heterocyclic compound) is shown on the bottom. Purines include adenine, guanine, hypoxanthine and xanthine, while pyrimidines include uracil, thymine and cytosine.

Figure A1 Martins et al.

**Purines**

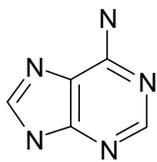 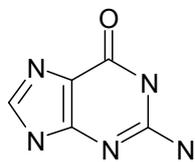 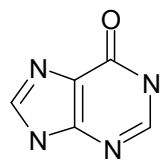 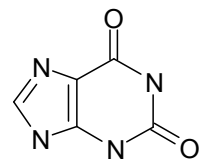

    Adenine               Guanine            Hypoxanthine        Xanthine

**Pyrimidines**

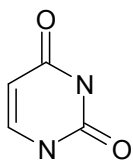 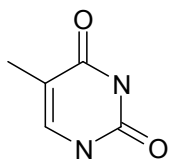 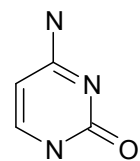

    Uracil               Thymine            Cytosine

**Table A1 -** Summary of the percentage of nucleobase standard material recovered after each separate cleaning step (charcoal filtration, hydrolysis and ion-exchange separation) during the purification process of the Murchison meteorite extract and soil. Values are not cumulative.

| Cleaning Step  Nucleobases | Charcoal cleaning | Hydrolysis | Ion-exchange separation |
|---|---|---|---|
| Cytosine | n.d. | 85 | 96 |
| Hypoxanthine | 97 | 95 | 69 |
| Guanine | 47 | 95 | 57 |
| Xanthine | 72 | 82 | 48 |
| Adenine | 53 | 83 | 78 |
| Uracil | 88 | 96 | 86 |
| Thymine | 69 | 81 | 68 |
| NR | 71 | 88 | 72 |

n.d. - not determined.
NR (nucleobase recovery) - This value corresponds to the percentage of the yields of nucleobases after each specific purification step.
The total nucleobase recovery (TNR) was calculated by considering that we applied the charcoal filtration step twice, the hydrolysis step once and the ion-exchange step once. TNR (in percentage) = 71% (charcoal filtration) x 88 % (hydrolysis) x 71% (charcoal cleaning) x 72 % (ion-exchange) = 32%